\documentstyle[preprint,aps]{revtex}
\tighten
\draft
\widetext
\input epsf
\preprint{HUTP-98/A061, NUB 3186}
\bigskip
\bigskip
\begin{document}
\title{Large $N$ Gauge Theories from Orientifolds with NS-NS $B$-flux}
\medskip
\author{Zurab Kakushadze\footnote{E-mail: 
zurab@string.harvard.edu}}
\bigskip
\address{Lyman Laboratory of Physics, Harvard University, Cambridge, 
MA 02138\\
and\\
Department of Physics, Northeastern University, Boston, MA 02115}
\date{August 9, 1998}
\bigskip
\medskip
\maketitle

\begin{abstract}
{}We consider D-branes and orientifold planes embedded in non-compact (orbifolded) space-time. We point out that even in the non-compact cases we can turn on non-zero (quantized) NS-NS antisymmetric $B$-field. In particular, we study the effect of the $B$-field
on four dimensional large $N$ gauge theories from orientifolds. Thus, in theories with both D3- and D7-branes, the effect of the $B$-field is non-trivial: the number of D7-branes (of each species) is reduced from 8 (which is the required number if the $B$-field is trivial) to 4. This results in a different orientifold string theory, and, subsequently, the corresponding large $N$ gauge theory is also different. We explicitly construct large $N$ gauge theories from orientifolds with non-zero $B$-field backgrounds with ${\cal N}=2,1,0$ supersymmetries. These theories, just as their counterparts without the $B$-field, have the property that in the large $N$ limit computation of any $M$-point correlation function reduces to the corresponding computation in the parent ${\cal N}=4$ supersymmetric theory with a unitary gauge group.

\end{abstract}
\pacs{}

\section{Introduction}

{}Recently large $N$ gauge theories have attracted a great deal of attention. This was
motivated by the AdS/CFT correspondence \cite{ads}. One of the implications of these developments was
a set of conjectures proposed in \cite{KaSi,LNV} which state that certain gauge theories with
${\cal N}=0,1$ supersymmetries are (super)conformal. These gauge theories are constructed by 
starting from a $U(N)$ gauge theory with ${\cal N}=4$ space-time supersymmetry in four dimensions, and orbifolding by a finite discrete subgroup $\Gamma$ of the $R$-symmetry group
$Spin(6)$ \cite{LNV}. These conjectures were shown at one-loop level for ${\cal N}=0$ theories
\cite{KaSi,LNV}, and to two loops for ${\cal N}=1$ theories using ordinary field theory techniques \cite{LNV}. 

{}In the subsequent development \cite{BKV} these conjectures were shown to be correct to
all loop orders in the large $N$ limit of 't Hooft \cite{thooft}.
The key observation in \cite{BKV} is the 
following. The above gauge theories can be obtained in the $\alpha^\prime\rightarrow 0$ limit
of Type IIB string theory with $N$ parallel D3-branes with the space transverse to the D-branes being ${\bf R}^6/\Gamma$. The gauge theory living in the world-volume of the D3-branes arises as a low energy effective field theory of {\em oriented} open strings that start and end on the D-branes.
The 't Hooft's large $N$ limit then corresponds to taking the limit $N\rightarrow \infty$ with
$\lambda=N\lambda_s$ fixed, where $\lambda_s$ is the Type IIB string coupling. In this context
a world-sheet with $g$ handles (corresponding to closed string loops) and $b$ boundaries (corresponding to D-branes) is weighted with
\begin{equation}\label{thoo}
 (N\lambda_s)^b  \lambda^{2g-2}_s=\lambda^{2g-2+b} N^{-2g+2}~.
\end{equation}  
With the identification $\lambda_s=g_{YM}^2$ we arrive at precisely the large $N$ expansion in the sense of 't Hooft (provided that the effective coupling $\lambda$ is fixed at a weak coupling value)\footnote{A similar expansion was discussed by Witten \cite{CS} for the case of three
dimensional Chern-Simons gauge theory where the boundaries of the string world-sheet are
``topological'' D-branes.}.

{}In \cite{BKV} the above idea was applied to prove that four dimensional 
gauge theories (including the cases with no supersymmetry) considered in
\cite{KaSi,LNV}
are conformal to all orders in perturbation theory in the large $N$ 
limit. The ultraviolet finiteness of string theory (that is, one-loop tadpole cancellation 
conditions) was shown to imply that the resulting (non-Abelian) gauge theories
where conformal in the large $N$ limit (in all loop orders). Moreover, in \cite{BKV}
it was also proven that computation of any correlation function in these theories 
in the large $N$ limit reduces to the corresponding computation in the parent 
${\cal N}=4$ supersymmetric gauge theory.

{}The work in \cite{BKV} was generalized in \cite{zura} where 
the setup was Type IIB string theory with D3-branes as well as orientifold planes
embedded in the orbifolded space-time. (In certain cases string consistency also requires presence of D7-branes.) This corresponds to Type IIB orientifolds. Introducing orientifold planes
is necessary to obtain $SO$ and $Sp$ gauge groups (without orientifold planes the gauge group
is always unitary), and also allows for additional variety in possible matter content. The presence of orientifold planes changes the possible topologies of the world-sheet.
Now we can have a world-sheet with $b$ boundaries, 
$c$ cross-caps (corresponding to orientifold planes), and $g$ handles. Such a 
world-sheet is weighted with     
\begin{equation}\label{thoo1}
 (N\lambda_s)^b \lambda_s^c \lambda^{2g-2}_s=\lambda^{2g-2+b+c} N^{-c-2g+2}~.
\end{equation}
Note that addition of a cross-cap results in a diagram suppressed by an additional 
power of $N$, so that in the large $N$ limit the cross-cap contributions are subleading.
In fact, in \cite{zura} it was shown that for string vacua which are perturbatively consistent 
(that is, the tadpoles cancel) calculations of correlation functions 
in ${\cal N}<4$ gauge theories
reduce to the corresponding calculations in the parent ${\cal N}=4$
{\em oriented} theory. This holds not only for finite (in the large $N$ limit) gauge theories
but also for the gauge theories which are not conformal. (In the latter case the gauge coupling running was shown to be suppressed in the large $N$ limit.) Here we note that the power of string perturbation techniques is an invaluable tool in proving such statements\footnote{In \cite{BJ}
the proofs of \cite{BKV} were cast into the field theory language. This is straightforward to do
once the string expansion is identified with 't Hooft's large $N$ expansion
as in \cite{BKV} in the case
of unitary gauge theories (that is, in absence of orientifold planes). However, 
string theory gives much more insight than the field theory approach ({\em e.g.}, it explains 
why the orbifold action on Chan-Paton matrices must be in an $n$-fold copy of the regular representation of the orbifold group). In the cases with 
orientifold planes string theory gives an insight for why the number of consistent theories of
this type is so limited, whereas within the field theory this fact seems to be a bit mysterious.}.

{}One distinguishing feature of large $N$ gauge theories obtained via
orientifolds is that the number
of possibilities which possess well defined world-sheet expansion (that is, are perturbative
from the orientifold viewpoint) is rather limited. This might appear surprising at the first sight, since naively one might expect that any choice of the orbifold group $\Gamma\subset Spin(6)$ should lead to a consistent world-sheet theory. As was recently discussed at length in \cite{KST}, in most cases
orientifolds contain non-perturbative sectors (arising from D-branes wrapping various collapsed two-cycles at orbifold singularities) which have no world-sheet description. 
In \cite{zura1} these conclusions were confirmed in the context of large $N$ gauge theories from orientifolds. This provides 
a very non-trivial check for correctness of the conclusions of \cite{KST}
concerning perturbative consistency of various compact orientifold models discussed in the
literature.

{}The setup of \cite{zura} corresponds to considering D-branes and orientifold planes embedded in a non-compact (orbifolded) space-time with zero NS-NS antisymmetric tensor backgrounds. Naively, it might (on the account of the space transverse to the D-branes being non-compact) appear that we cannot consider non-trivial $B$-field backgrounds. This is not quite so, however. In particular, the subject of this paper is precisely to study the effects of non-zero $B$-field backgrounds on large $N$ gauge theories from orientifolds. Thus, we point out that even in the non-compact cases we can turn on non-zero quantized $B$-field which in certain cases has a non-trivial effect. More precisely, if the $B$-field is turned on in the directions within the world-volume of the D-branes, in the non-compact limit it has no effect on the world-volume theory. However, if the $B$-field is turned on in the directions transverse to the D-branes, it has the effect of reducing the corresponding tadpoles by a factor of 2. More concretely, the number of D-branes required to cancel the corresponding R-R charge is reduced by 2. Since the number of D3-branes in such theories is arbitrary, the effect of the $B$-field on D3-branes is immaterial. However, if we consider theories with D3- as well as D7-branes, we have a non-trivial modification due to the $B$-field: the number of D7-branes (of each species) is reduced from 8 (which is the required number if the $B$-field is trivial) to 4. This results in a different orientifold string theory, and, subsequently, the corresponding large $N$ gauge theory is also different. In this paper we explicitly construct large $N$ gauge theories from orientifolds with non-zero $B$-field backgrounds with ${\cal N}=2,1,0$ supersymmetries. These theories, just as their counterparts without the $B$-field, have the property that in the large $N$ limit computation of any $M$-point correlation function reduces to the corresponding computation in the parent ${\cal N}=4$ supersymmetric theory.

{}The remainder of this paper is organized as follows. In section II we review the setup of \cite{zura}, and also its generalization \cite{zura2} which allows to include non-supersymmetric theories. We also discuss the effect of the $B$-field on the D-brane theories.
In sections III, IV, V, and VI we construct various large $N$ gauge theories from orientifolds with ${\cal N}=2,1,0$ supersymmetries. In section VII we discuss various issues relevant for the previous discussions. In particular, we point out how the effect of the $B$-field can be understood from the AdS viewpoint in the context of F-theory.

\section{Preliminaries}

{}{}In this section we review the setup in \cite{zura} which leads to supersymmetric 
large $N$ gauge theories from orientifolds. 
We also briefly review a simple generalization discussed in \cite{zura2} that allows to 
consider non-supersymmetric cases as well. We then discuss the effects of turning on a non-zero $B$-field which lead to construction of new large $N$ gauge theories from orientifolds discussed in this paper.

\subsection{Setup}

{}Consider Type IIB string theory on ${\bf C}^3/\Gamma$ where
$\Gamma\subset SU(3)(SU(2))$ so that the resulting theory has ${\cal N}=2(4)$ 
supersymmetry in four dimensions. In the following we will use the following notations: $\Gamma=\{g_a\vert a=1,\dots,|\Gamma|\}$
($g_1=1$).
Consider the $\Omega J$ orientifold of this 
theory, where $\Omega$ is the world-sheet parity reversal, and $J$ 
is a ${\bf Z}_2$ element ($J^2=1$) acting on the complex coordinates $z_i$
($i=1,2,3$) on ${\bf C}^3$ as follows: $J z_i=-Jz_i$. The resulting theory has ${\cal N}=1(2)$
supersymmetry in four dimensions. 
 
{}Note that we have an orientifold 3-plane corresponding to the $\Omega J$
element of the orientifold group. If $\Gamma$ has
a ${\bf Z}_2$ subgroup, then we also have an orientifold 7-plane.
If we have an orientifold 7-plane we must 
introduce 8 of the corresponding D7-branes to cancel the R-R charge appropriately.
(The number 8 of D7-branes is required by the corresponding tadpole cancellation
conditions.) Note, however, that the number of D3-branes is not constrained (for the corresponding untwisted tadpoles automatically vanish in the non-compact case).

{}We need to specify the action of $\Gamma$ on the Chan-Paton factors
corresponding to the D3- and D7-branes.  
These are given by Chan-Paton matrices which we collectively refer to
as $n^\mu \times n^\mu$ matrices $\gamma^\mu_a$, where the superscript $\mu$ refers to the corresponding
D3- or D7-branes. Note that ${\mbox{Tr}}(\gamma^\mu_1)=n^\mu$ where 
$n^\mu$ is the number of D-branes labelled by $\mu$. 

{}At one-loop level there are three different sources for massless tadpoles:
the Klein bottle, annulus, and M{\"o}bius strip amplitudes. The factorization property of string theory implies that the tadpole cancellation conditions read (see, {\em e.g.}, \cite{zura,rev}
for a more detailed discussion):
\begin{equation}\label{BC}
 B_a+\sum_\mu C^\mu_a {\mbox{Tr}}(\gamma^\mu_a)=0~.
\end{equation}
Here $B_a$ and $C^\mu_a$ are (model dependent) numerical coefficients of order 1. 

{}In the world-volume of D3-branes there lives a four dimensional ${\cal N}=1(2)$ supersymmetric gauge theory (which is obtained in the low energy, that is, $\alpha^\prime\rightarrow 0$ limit). Since the number of D3-branes is unconstrained, we can consider the large
$N$ limit of this gauge theory. In \cite{zura} (generalizing the work in \cite{BKV}) it was shown that, if for a given
choice of the orbifold group $\Gamma$ the world-sheet description for the orientifold is adequate, 
then in the large $N$ limit (with $\lambda=N\lambda_s$ fixed, where $\lambda_s$ is the Type IIB string coupling) computation of any correlation function in this gauge theory is reduced to the corresponding computation in the parent ${\cal N}=4$ supersymmetric {\em oriented} gauge theory before orbifolding and orientifolding. In particular, the 
running of the gauge coupling is suppressed in the large $N$ limit. Moreover, if \begin{equation}\label{Klein}
 {\mbox {Tr}}(\gamma^\mu_a)=0~\forall a\not=1
\end{equation}
(that is, $B_a=0$ $\forall a\not=1$), then the the one-loop $\beta$-function coefficients $b_0$
for non-Abelian gauge theories living in world-volumes of the D3-branes vanish.

{}Generically, however, twisted Chan-Paton matrices are not traceless, so that 
the condition (\ref{Klein}) is not satisfied. This generally leads to gauge theories with non-vanishing one-loop $\beta$-functions. Nonetheless, 
``non-finiteness'' of such theories is a subleading 
effect in the large $N$ limit. This is because the $\beta$-function coefficients grow
as
\begin{equation}\label{beta}
 b_s=O(N^s)~,~~~s=0,1,\dots~,
\end{equation}
instead of $b_s=O(N^{s+1})$ (as in, say, pure $SU(N)$ gauge theory) \cite{zura}.
These estimates for the $\beta$-function coefficients for $b_{s>0}$
are non-trivial from the field theory point of view as they imply non-trivial cancellations
between couplings (such as Yukawas) in the gauge theory. On the other hand, within 
string perturbation expansion these statements become obvious once we carefully 
consider twisted boundary conditions and tadpole cancellation \cite{BKV,zura}.

\subsection{Discrete Torsion and Non-Supersymmetric Models}

{}One can generalize the above discussion to include non-supersymmetric large $N$ gauge theories from orientifolds \cite{zura2}. The main subtlety that arises here is the presence of tachyons in the twisted closed string sectors. (Note that there are no tachyons present in the open string sectors.) As we discuss in a moment, the presence of tachyons by itself does not pose a problem for the consistency of the corresponding large $N$ theories. However, if tachyons are present in the physical closed string spectrum of the corresponding orientifold model, {\em a priori} they too contribute into the tadpoles. Moreover, the cancellation conditions for the tachyonic and massless tadpoles generically are rather different. That is, the numerical coefficients $B_a$ and $C_a$ in (\ref{BC}) corresponding to the massless and tachyonic tadpoles are generically different. This typically overconstrains the tadpole cancellation conditions, which makes it rather difficult to find tadpole free non-supersymmetric orientifolds. 

{}There is a way around the above difficulties, however. Let $\Gamma\subset SU(3)$ be an orbifold group such that it contains a ${\bf Z}_2$ subgroup. Let the generator of this ${\bf Z}_2$ subgroup be $R$. Consider now the following orbifold group: $\Gamma^\prime=\{g_a^\prime| 
a=1,\dots,|\Gamma|\}$, 
where the elements $g^\prime_a$ are the same as $g_a$ except that $R$
is replaced everywhere by $R^\prime=R T$, where $T$ is the generator of a ${\bf Z}_2$ group 
corresponding to the {\em discrete torsion}. The action of $T$ is defined as follows: it acts as identity in the bosonic sectors (that is, in the NS-NS and R-R closed string sectors, and in the NS open string sector), and it acts as $-1$ in the fermionic sectors
(that is, in the NS-R and R-NS closed string sectors, and in the R open string sector) when, say, acting on the ground states. Now consider Type IIB on ${\bf C}^3/\Gamma^\prime$. Generically, in this theory all supersymmetries are broken. 
There are certain ``exceptions'', however. Thus,
if $\Gamma\approx{\bf Z}_2$ such that $\Gamma\subset SU(2)$, then inclusion of the discrete torsion does not break supersymmetry. Similarly, if $\Gamma\approx{\bf Z}_2\otimes {\bf Z}_2$ such that $\Gamma\subset SU(3)$, the number of unbroken supersymmetries is not affected by the discrete torsion. The basic reason for this is that the ${\bf Z}_2$ twist is self-conjugate. On the other hand, in certain cases we can include the discrete torsion in ways slightly different from the one just described. In particular, let $\Gamma\approx{\bf Z}_4$ such that $\Gamma\subset SU(2)$. Let $g$ be the generator of this ${\bf Z}_4$. Next, consider the following orbifold group:
$\Gamma^\prime=\{1,gT,g^2,g^3 T\}$. In other words, the ${\bf Z}_4$ twists $g$ and $g^3$ (but not the corresponding ${\bf Z}_2$ twist $g^2$) are accompanied by the discrete torsion $T$. In this case we also have no unbroken supersymmetries. More generally, we can include the discrete torsion if the orbifold group $\Gamma$ contains a ${\bf Z}_{2^n}$ subgroup. However, the corresponding orbifold group $\Gamma^\prime$ does not always lead to non-supersymmetric theories.    

{}Next, suppose using the above construction we have found an orbifold group $\Gamma^\prime$ such that Type IIB on ${\bf C}^3/\Gamma^\prime$ is non-supersymmetric. Then it is not difficult to show that the following statement holds. If $\Gamma$ is such that the $\Omega J$ orientifold of Type IIB on ${\bf  C}^3/\Gamma$ is a perturbatively well defined 
${\cal N}=1$ or ${\cal N}=2$ theory (that is, all the massless tadpoles cancel, and there are no non-perturbative contributions to the massless spectrum), then in the $\Omega J$ orientifold of Type IIB on ${\bf  C}^3/\Gamma^\prime$ (which is non-supersymmetric) all the tachyonic and massless one-loop tadpoles (and, consequently, all the anomalies) automatically cancel. 
This is basically due to the fact that in all perturbatively well defined orientifolds the twisted Chan-Paton matrices corresponding to the sectors which become tachyonic upon adding the discrete torsion $T$ are all traceless (in both bosonic and fermionic open loop channels, or, equivalently, in both NS-NS and R-R closed tree channels). For this reason the tachyonic tadpoles are always multiplied by zero, and, therefore, cancel.
This enables one to construct anomaly free non-supersymmetric large $N$ gauge theories from orientifolds \cite{zura2}. In particular, in this paper we will generalize the construction of \cite{zura2} (along with the construction of supersymmetric theories in \cite{zura,zura1}) to include non-trivial $B$-field backgrounds.

{}Next, we would like to comment on the consistency of non-supersymmetric theories.
This is an important point since they contain tachyons. However, 
here we are interested in large $N$ gauge theories in the 't Hooft limit $N\lambda_s={\mbox{fixed}}$, which implies that the closed string coupling constant $\lambda_s\rightarrow 0$ as we take $N$ to infinity. In particular, all the world-sheets with handles (corresponding to closed string loops) as well as cross-caps (corresponding to orientifold planes) are suppressed in this limit. That is, the closed string sector decouples from the open string sector in this limit, and after taking $\alpha^\prime\rightarrow 0$ we can ignore the closed sting states (regardless of whether they are tachyonic or not) altogether. In other words, the string construction here is simply an efficient and fast way of obtaining a field theory result, and at the end of the day we are going to throw out all the irrelevant ingredients (such as complications in the closed string sectors due to the presence of tachyons) and keep only those relevant for the field theory discussion. Note that we would not be able to do the same had we considered a {\em compact} model (with a finite number of D3-branes). In this case the closed string sector does not decouple and the theory is sick due to the presence of tachyons\footnote{In certain cases, however, it is plausible that one might still be able to maintain at least some control over the vacuum structure.}. 

\subsection{Perturbative Orientifolds}

{}The arguments of \cite{zura} that imply the above properties of D3-brane gauge theories
are intrinsically perturbative. In particular, a consistent world-sheet expansion is crucial for their
validity. It is therefore important to understand the conditions for the perturbative orientifold
description to be adequate. 

{}Naively, one might expect that any choice of the orbifold group $\Gamma\subset Spin(6)$
(note that $Spin(6)$ is the $R$-symmetry group of ${\cal N}=4$ gauge theory)
should lead to an orientifold with well defined world-sheet expansion in terms of boundaries
(corresponding to D-branes), cross-caps (corresponding to orientifold planes) and handles
(corresponding to closed string loops). This is, however, not the case \cite{KST,zura1,class}. 
In fact, the number of choices of $\Gamma$ for which such a world-sheet expansion is adequate is rather constrained. In particular, in \cite{KST,zura1,class} it was argued that for ${\cal N}=1$ 
there are only seven choices of the orbifold group leading to consistent perturbative orientifolds: 
${\bf Z}_2\otimes {\bf Z}_2$, ${\bf Z}_3$, ${\bf Z}_7$, ${\bf Z}_3\otimes {\bf Z}_3$, ${\bf Z}_6$ , ${\bf Z}_2\otimes{\bf Z}_2\otimes {\bf Z}_3$ \cite{KST,zura1}, and
$\Delta(3\cdot 3^2)$ (the latter group is non-Abelian) \cite {class}. All the other orbifold groups
lead to orientifolds containing sectors which are non-perturbative from the orientifold viewpoint (that is, these sectors have no world-sheet description). These sectors can be thought of as arising from D-branes wrapping various (collapsed) 2-cycles in the orbifold. 

{}As to the ${\cal N}=2$ theories, in \cite{zura} it was shown that the corresponding orientifolds
are perturbatively well defined only for the following choices of the orbifold group:
${\bf Z}_M$, $M=2,3,4,6$. Non-supersymmetric orientifolds with non-chiral spectra can be obtained by starting from the ${\cal N}=2$ orbifolds ${\bf Z}_4$ and ${\bf Z}_6$ and including the
discrete torsion $T$ as discussed in the previous subsection \cite{zura2}. The resulting orientifolds are perturbatively well defined. To obtain chiral models, one starts from the ${\cal N}=1$ orbifolds ${\bf Z}_6$ and ${\bf Z}_2\otimes{\bf Z}_2\otimes {\bf Z}_3$, and includes the discrete torsion $T$. These cases also appear to be perturbatively well defined, although the corresponding arguments are less robust than in other cases (see subsection C of section VI for more details). The above restrictions on the orbifold group will be important in the construction of consistent large $N$ gauge theories from orientifolds with the $B$-flux. 

\subsection{Effects of the $B$-flux}

{}In this subsection we discuss the key effects which result by turning on a non-zero $B$-field.
Note that the untwisted NS-NS two-form $B_{\mu\nu}$ is
odd under the orientifold projection $\Omega$. This implies that the corresponding states are
projected out of the closed string massless spectrum. So, naively, it might appear that we cannot have any non-trivial $B$-flux. This is, however, not the case. Thus, consider for a moment compactifying some of the directions transverse to D3-branes. If we compactify more than one direction, then we can have a non-zero $B$-flux in those directions. To see this recall that in the compact directions $B_{ij}$ is defined up to a unit shift: $B_{ij}\sim B_{ij}+1$. With this normalization,
the only two values of $B_{ij}$ invariant under $\Omega$ are 0 and $1/2$, hence quantization of
$B_{ij}$. Now we can consider the decompactification limit. In this limit we can have distinct vacua with $B_{ij}=0$ and $B_{ij}\not=0$. More generally we can also consider D7-branes. We can have vacua with non-zero $B$-field (in the decompactification limit) which is either in the directions transverse to the D7-branes or within their world-volumes. As we discuss in a moment, 
there is a key difference between these two cases.  

{}Thus, consider D7-branes in ${\bf R}^{10}$. More precisely, we need to add orientifold 7-planes to cancel the R-R charge appropriately. Now consider compactifying two of the {\em space-like} directions {\em along} the D7-branes, call them $X^1$ and $X^2$, on $T^2\approx S^1\otimes S^1$,
where the radii of the two circles are $r_1$ and $r_2$, respectively. 
Let us consider the case where $B_{12}=0$.
In the closed string sector the left- and right-moving momenta are given by ($a=1,2$):
\begin{equation}
 P^a_{L,R} = {m^a\over 2r_a}\pm n^a r_a~,
\end{equation} 
where $m^a,n^a\in {\bf Z}$ are the momentum and winding numbers, respectively. Next, consider the following freely acting orbifold of this theory.
Let $S_a$ be a half winding shift in the $X^a$ direction. In other words, the action of $S_a$ on the left- and right-moving momenta is given by:
\begin{equation}\label{twist}
 S_a|P^a_L,P^a_R\rangle = \exp\left(\pi i (P^a_L+P^a_R) r_a\right)|P^a_L,P^a_R\rangle~.
\end{equation}  
In the untwisted sectors we have:
\begin{equation}
 S_a|P^a_L,P^a_R\rangle = \exp\left(\pi i m_a \right)|P^a_L,P^a_R\rangle~.
\end{equation}
Thus, the states with even momentum numbers $m_a$ are invariant under these twists, whereas
for odd momentum number states we have corresponding $-1$ phases under $S_1$, $S_2$, or both. 

{}Modular invariance of the (oriented) closed string sector requires that we add twisted sectors
labeled by $S_1$, $S_2$ and $S_1S_2$. The corresponding left- and right-moving momenta read: 
\begin{equation}
 P^a_{L,R} (\alpha^1,\alpha^2)= {m^a\over 2r_a}\pm (n^a+{1\over 2}\alpha^a) r_a~,
\end{equation}
where $(\alpha^1,\alpha^2)=(1,0)$, $(\alpha^1,\alpha^2)=(0,1)$ and $(\alpha^1,\alpha^2)=(1,1)$
in the $S_1$, $S_2$ and $S_1S_2$ twisted sectors, respectively. Thus, we can use the same formula in the untwisted sector as well which is now labeled by $(\alpha^1,\alpha^2)=(0,0)$.
We have to define the action of the twists $S_1$ and $S_2$ on the twisted sector states. There are two inequivalent consistent choices here. First, we can have the same action in the twisted sectors as in the untwisted sector, that is, the action given by (\ref{twist}) (where we replace
$P^a_{L,R}$ by $P^a_{L,R} (\alpha^1,\alpha^2)$). In this case it is not difficult to show that the resulting theory is the same as a compactification of a new torus, call it 
${\widetilde T}^2$, which factorizes into two circles with radii ${\widetilde r}^a=r^a/2$. In other words, the ${\bf Z}_2\otimes {\bf Z}_2$ freely acting orbifold just results in reducing the radii of the circles by a factor of 2. As to the $B$-field, in this case it is still zero after orbifolding.

{}The second possibility is that $S_1$ acting in the $S_2$ twisted sector has an extra minus sign. Then the consistency requires that $S_2$ acting in the $S_1$ twisted sector also has an extra minus sign. Moreover both $S_1$ and $S_2$ acting in the $S_1 S_2$ twisted sector also have an extra minus sign. These extra phases are nothing but a discrete torsion. It is not difficult to show that the resulting theory is the same as a compactification on the above ${\widetilde T}^2$ except that there is a non-zero $B$-field now:
$B_{12}=1/2$. 

{}Let us now consider the action in the open string sector. It is described by the Chan-Paton matrices $\gamma_{S_a}$ which must form a (projective) representation of ${\bf Z}_2\otimes {\bf Z}_2$. Here {\em a priori} we have two choices:
\begin{equation}
 \gamma_{S_1}\gamma_{S_2} = \epsilon \gamma_{S_2}\gamma_{S_1}~,
\end{equation} 
$\epsilon=\pm1$. In the case without the discrete torsion between $S_1$ and $S_2$ we must choose $\epsilon=+1$ since the orbifold action in the closed string sector is equivalent to simply rescaling the radii. In the open string sector the action of the {\em commuting} Chan-Paton matrices $\gamma_{S_1}$ and $\gamma_{S_2}$ can be non-trivial. In fact, it is arbitrary up to the requirement that $\epsilon=+1$. In particular, there are no tadpoles for any choice of these matrices. Thus, the effect of this freely acting orbifold in the open string sector reduces to having two discrete Wilson lines (which commute with each other). 

{}Now let us consider the case with the discrete torsion, {\em i.e.}, the case with non-zero $B$-field. Then the string consistency requires that we have $\epsilon=-1$ \cite{Bij,wit,NS}.
Note that here we can also view the action or the orbifold as having two Wilson lines except that now they do not commute. This results in reduction of the rank of the gauge group coming from the D7-branes by a factor of 2. In particular, half of the original Cartan generators are projected out of the massless spectrum. More precisely, the lightest states with the corresponding quantum numbers have non-zero momentum numbers (but the winding numbers can be zero). 

{}This is the key point for the following discussion. Thus, let us take the decompactification limit where $r_1,r_2\rightarrow\infty$. Then all of these states become massless, and the original gauge symmetry is restored, {\em i.e.}, no rank reduction occurs in this limit. The upshot here is that in the decompactification limit the D-branes do {\em not} feel the presence of the quantized $B$-field if it is turned on in the directions within the world-volume of the D-branes.

{}Now let us consider the case where we compactify on $T^2\approx S^1\otimes S^1$ two space-like directions {\em transverse} to D7-branes. The situation here is similar to that discussed above except that the roles of the windings and the momenta are interchanged. This makes all the difference. Thus, we can now consider a freely acting orbifold but this time we must shift by half of the momenta. For the case of the discrete torsion, which corresponds to turning on half-integer $B$-field, we again have rank reduction, but this time the lightest states
with the quantum numbers of the states projected out of the massless spectrum have non-zero winding numbers. In the decompactification limit these states, therefore, become infinitely heavy and decouple. The net result is rank reduction by a factor of 2. Alternatively, we can view this as having only half of the D-branes compared with the case without the $B$-field. (More precisely, the D-branes split into two equal groups, and only one linear combination is kept.)   

{}Let us summarize the above discussion. If we have a non-zero $B$-field in the directions 
within the world-volume of the D-branes, then it has no effect on the D-brane theory in the non-compact case. If the $B$-field is in the directions transverse to the D-branes, then we have half as many D-branes than in the case without the $B$-field. Note that any non-trivial $B$-field transverse to the D3-branes is going to reduce the rank of the corresponding gauge group. However, the number of D3-branes in the theories we are interested in here is unlimited, so the net effect is immaterial. Thus, the only theories that we are going to consider are those with D7-branes. Without the $B$-field their number is 8 (for each set of D7-branes - see below). However, if we have non-zero $B$-field in the directions transverse to the D7-branes, we have only 4 of them. This changes the resulting large $N$ gauge theory.

{}Let $z_i$, $i=1,2,3$, be the complex coordinates parametrizing the space transverse to D3-branes. Let D$7_i$-branes be located at points in the $z_i$ complex plane. Also, let $b_i=0,1/2$ be the values of the $B$-field in the directions corresponding to $z_i$. Then the number of D$7_i$-branes (if any) is given by $n_i=8/k_i$, where $k_i=2^{2b_i}$. Moreover,
the $7_i7_j$ ($i\not=j$) open string sector states come with multiplicity $k_i k_j$. This is due to the discrete gauge symmetry (isomorphic to ${\bf Z}_2$ for each each D$7_i$-brane with $k_i=2$) which arises in the process of the rank reduction \cite{NS,class}. Similarly, the $37_i$ open string sector states come with multiplicity $k_i$.   

{}Another important point is that some of the twisted tadpole cancellation conditions get modified
for non-zero $B$-field configurations \cite{NS,class}. Suppose $g_a$ is an element of the orbifold group $\Gamma$ such that the fixed point locus of the corresponding twist is of real dimension two. Then, if the components of $B_{ij}$ corresponding to this locus form a non-zero $2\times 2$ matrix (that is, an antisymmetric matrix of rank 2), 
some of the coefficients $B_a$ in (\ref{BC}) are modified. More precisely, if $g_a$ is the generator of a ${\bf Z}_M$ subgroup of the orbifold group $\Gamma$ such that $M\in 2{\bf Z}+1$ (in this paper these are the cases with a ${\bf Z}_3$ subgroup), 
then the sign of the coefficient $B_a$ is flipped. 
None of the other twisted tadpoles get modified. Thus, following these rules
one can read off the twisted tadpole cancellation conditions for the cases with the $B$-field 
from the corresponding tadpole cancellation conditions for backgrounds with the
trivial $B$-field.

\section{${\cal N}=2$ Gauge Theories}

{}In this section we construct four dimensional ${\cal N}=2$ gauge 
theories form Type IIB orientifolds with the $B$-field. We start from Type IIB string theory on ${\cal M}$=
${\bf C}^3/\Gamma$, where $\Gamma=\{g^k\vert k=0,\dots,M-1\}\approx {\bf Z}_M$
($M=2,4,6$) 
is the orbifold group whose action on the complex coordinates $z_i$ ($i=1,2,3$) 
on ${\bf C}^3$ is given by $gz_1=z_1$, $gz_2=\omega z_2$, $gz_3=\omega^{-1} z_3$ 
($\omega=\exp(2\pi i/M)$). The $B$-field is turned on in the 
directions corresponding to $z_1$. 
Next, we consider the $\Omega J$ orientifolds of these theories.

{}The number of D3-branes is arbitrary, whereas the number of D7-branes (whose locations are given by points in the $z_1$ complex plane) is 4. The twisted tadpole cancellation conditions have the following solutions:
\\
$\bullet$ $M=2$ ($N=n_3/2$):
\begin{eqnarray}
 &&\gamma_{1,3}={\mbox{diag}}(i~(N~{\mbox{times}}), 
 -i~(N~{\mbox{times}}))~,\\
 &&\gamma_{1,7}={\mbox{diag}}(i~(2~{\mbox{times}}), -i~(2~{\mbox{times}}))~.
\end{eqnarray}
$\bullet$ $M=4$ ($N=n_3/4$):
\begin{eqnarray}
 \gamma_{1,3}={\mbox{diag}}(&&\exp(\pi i/4)~(N~{\mbox{times}}),
 \exp(-\pi i/4)~(N~{\mbox{times}}),\nonumber\\
 &&\exp(3\pi i/4)~(N~{\mbox{times}}),
 \exp(-3\pi i/4)~(N~{\mbox{times}}))~,\\
 \gamma_{1,7}={\mbox{diag}}(&&\exp(\pi i/4),
 \exp(-\pi i/4), \exp(3\pi i/4), \exp(-3\pi i/4))~.
\end{eqnarray}
$\bullet$ $M=6$ ($N=(n_3-2)/6$):
\begin{eqnarray}
 \gamma_{1,3}={\mbox{diag}}(&&
 i\exp(2\pi i/3)~(N~{\mbox{times}}),
 -i\exp(2\pi i/3)~(N~{\mbox{times}}),\nonumber\\
 &&i\exp(-2\pi i/3)~(N~{\mbox{times}}),
 -i\exp(-2\pi i/3)~(N~{\mbox{times}})),\nonumber\\
 &&i~((N-1)~{\mbox{times}}),
 -i~((N-1)~{\mbox{times}}))~,\\ 
 \gamma_{1,7}={\mbox{diag}}(&&
 i\exp(2\pi i/3),-i\exp(2\pi i/3),i\exp(-2\pi i/3),-i\exp(-2\pi i/3))~.
\end{eqnarray}
The massless open string spectra of these models is given in Table \ref{spectrum1}. 

{}Note that the one-loop $\beta$-function coefficients for the non-Abelian subgroups of the 33 sector gauge group vanish in the ${\bf Z}_2$ and ${\bf Z}_4$ theories, whereas in the ${\bf Z}_6$ theory we have:
\begin{eqnarray}
 && b_0 (N)=+1~,\\
 && b_0 (N-1)=-2~,
\end{eqnarray}
where $b_0 (N)$ and $b_0 (N-1)$ are the one-loop $\beta$-function coefficients for the $SU(N)$ and $SU(N-1)$ subgroups of the 33 sector gauge group.

{}Here the following remark is in order. Note that in the ${\bf Z}_6$ case the one-loop $\beta$-function coefficients of the non-Abelian subgroups of the 33 open string sector gauge group are non-zero, whereas in the ${\bf Z}_2$ and ${\bf Z}_4$ cases they vanish. This is in accord with the observation of \cite{zura} that $b_0=0$ if all the twisted Chan-Paton matrices $\gamma_a$ ($a\not=1$) are traceless. Moreover, generically we do not expect $b_0$ coefficients to vanish unless all ${\mbox{Tr}}(\gamma_a)$ ($a\not=1$) are traceless. However, there can be ``accidental'' cancellations in some models such that all $b_0=0$ despite some of the twisted Chan-Paton matrices not being traceless. Such ``accidental'' cancellations, in particular, occur in the ${\cal N}=2$ supersymmetric ${\bf Z}_3$ and ${\bf Z}_6$ models 
{\em without} the $B$-field
discussed in \cite{zura}. These cancellations were explained in \cite{zura} using the results obtained in \cite{NS}. On the other hand, such an ``accidental'' cancellation does not occur in the above ${\bf Z}_6$ model with the $B$-field.

\section{${\cal N}=1$ Gauge Theories}

{}In this section we construct four dimensional ${\cal N}=1$ gauge 
theories form Type IIB orientifolds with the $B$-field.
We start from Type IIB string theory on ${\cal M}=
{\bf C}^3/\Gamma$, where $\Gamma\approx {\bf Z}_2\otimes 
{\bf Z}_2$, ${\bf Z}_2\otimes 
{\bf Z}_2\otimes {\bf Z}_3$, and ${\bf Z}_6$. 
Next, we consider the $\Omega J$ orientifolds of these theories.

\subsection{The ${\bf Z}_2\otimes {\bf Z}_2$ Models}

{}Consider the case where $\Gamma=\{1,R_1,R_2,R_3\}\approx {\bf Z}_2
\otimes {\bf Z}_2$ ($R_i R_j=R_k$, $i\not=j\not=k\not=i$) is the
orbifold group whose action on the complex coordinates $z_i$
is given by $R_i z_j=-(-1)^{\delta_{ij}} z_j$. 
Let $b_i=0,1/2$ be the values of the $B$-field in the directions corresponding to $z_i$. 
The number of the D3-branes is arbitrary. We have three sets of D7-branes which we will refer to as D$7_i$-branes. The locations of the D$7_i$-branes are given by the points in the $z_i$ complex plane. The number of D$7_i$-branes is $n_i=8/2^{2b_i}$. 

{}The twisted
tadpole cancellation conditions imply that the corresponding Chan-Paton matrices
$\gamma_{R_i,3}$ and $\gamma_{R_i,7_j}$ are traceless:
\begin{equation}
 {\mbox{Tr}}(\gamma_{R_i,3})={\mbox{Tr}}(\gamma_{R_i,7_j})=0~.
\end{equation} 
Up to equivalent representations, the solution to the twisted tadpole cancellation conditions is given by ($N=n_3/2$):
\begin{eqnarray}
 \gamma_{R_i,3}=i\sigma_i\otimes {\bf I}_N~,
\end{eqnarray} 
where $\sigma_i$ are Pauli matrices, and ${\bf I}_N$ is an $N\times N$ identity matrix.
(The action on the D$7_i$ Chan-Paton charges
is similar.) The spectra of these models are given in Table \ref{Z2}. Note that the one-loop $\beta$-function coefficient vanishes for the $Sp(N)$ 33 sector gauge group.

\subsection{The ${\bf Z}_2\otimes {\bf Z}_2\otimes {\bf Z}_3$ Models}

{}Let us now consider the case with $\Gamma
\approx {\bf Z}_2\otimes {\bf Z}_2\otimes {\bf Z}_3$. Let
$R_i$ be the same as in the previous example, and let $g$ be the generator of   
${\bf Z}_3\subset \Gamma$ which acts as as follows:
$gz_i=\omega z_i$, where $\omega=\exp(2\pi i/3)$.
Just as in the previous case, 
the untwisted tadpole cancellation conditions require presence of three sets of 
D7-branes with $n_i=8/2^{2b_i}$ D7-branes in each set. The solution to the twisted tadpole cancellation conditions reads ($N=(n_3+4)/6$):
\begin{eqnarray}
 &&\gamma_{g,3}={\mbox{diag}}(\omega {\bf I}_{2N},\omega^2 {\bf I}_{2N},{\bf I}_{2N-4})~,\\
 &&\gamma_{R_i,3}=i\sigma_i\otimes {\bf I}_{3N-2}~.
\end{eqnarray}
The action on D$7_i$ Chan-Paton charges is similar. 
The spectra of these models are given in Table \ref{Z2}. The non-Abelian gauge
anomaly is cancelled in these models. Also, the one-loop $\beta$-function coefficients for the $SU(N)$ and $Sp(N-2)$ subgroups of the 33 sector gauge group are given by:
\begin{eqnarray}
 && b_0 (N)=+3~,\\
 && b_0 (N-1)=-6~.
\end{eqnarray} 

\subsection{The ${\bf Z}_6$ Model}

{}Consider the case $\Gamma\approx{\bf Z}_6\approx{\bf Z}_2\otimes {\bf Z}_3$, where the generators $R$ and $g$ of the ${\bf Z}_2$ and ${\bf Z}_3$ subgroups have the following action on the complex coordinates $z_i$: $Rz_1=-z_1$, $Rz_2=-z_2$, $Rz_3=z_3$, $gz_i=\omega z_i$ ($\omega=\exp(2\pi i/3)$). The $B$-field is turned on in the directions corresponding to $z_3$. 
The number of the D3-branes is arbitrary, whereas the number of D7-branes (whose locations are given by points in the $z_3$ complex plane) is 4. The twisted tadpole cancellation conditions have the following solution ($N=(n_3+4)/6$):
\begin{eqnarray}
 &&\gamma_{g,3}={\mbox{diag}}(\omega {\bf I}_{2N},\omega^2 {\bf I}_{2N},{\bf I}_{2N-4})~,\\
 &&\gamma_{R,3}={\mbox{diag}}(i,-i)\otimes {\bf I}_{3N-2}~,\\
 &&\gamma_{g,7}={\mbox{diag}}(\omega {\bf I}_{2},\omega^2 {\bf I}_{2}) ~,\\
 &&\gamma_{R,7}={\mbox{diag}}(i,-i)\otimes {\bf I}_{2}~.
\end{eqnarray}
The spectrum of this model is given in Table \ref{Z2}. Note that the non-Abelian gauge anomaly cancels in this model. Also, the one-loop $\beta$-function coefficients for the $SU(N)$ and $SU(N-2)$ subgroups of the 33 sector gauge group are given by:
\begin{eqnarray}
 && b_0 (N)=+4~,\\
 && b_0 (N-1)=-8~.
\end{eqnarray}

\section{Non-Chiral ${\cal N}=0$ Gauge Theories}

{}In this section we construct non-chiral four dimensional ${\cal N}=0$ gauge 
theories form Type IIB orientifolds with the $B$-field. We start from Type IIB string theory on ${\cal M}$=
${\bf C}^3/\Gamma$, where $\Gamma=\{{\widetilde g}^k\vert k=0,\dots,M-1\}\approx {\bf Z}_M$
($M=4,6$) 
is the orbifold group whose action on the complex coordinates $z_i$ ($i=1,2,3$) 
on ${\bf C}^3$ is given by $gz_1=z_1$, $gz_2=\omega z_2$, $gz_3=\omega^{-1} z_3$ 
($\omega=\exp(2\pi i/M)$), and ${\widetilde g}=g T$. Here the action of $T$ is defined as follows: it acts as identity in the bosonic sectors (that is, in the NS-NS and R-R closed string sectors, and in the NS open string sector), and it acts as $-1$ in the fermionic sectors
(that is, in the NS-R and R-NS closed string sectors, and in the R open string sector) when, say, acting on the ground states. Note that including the discrete torsion $T$ in the orbifold action breaks all supersymmetries. The $B$-field is turned on in the 
directions corresponding to $z_1$. 
Next, we consider the $\Omega J$ orientifolds of this theories.

{}The number of D3-branes is arbitrary, whereas the number of D7-branes (whose locations are given by points in the $z_1$ complex plane) is 4. The twisted tadpole cancellation conditions have the following solutions:\\
$\bullet$ $M=4$ ($N=n_3/4$):
\begin{eqnarray}
 \gamma_{1,3}={\mbox{diag}}(&&\exp(\pi i/4)~(N~{\mbox{times}}),
 \exp(-\pi i/4)~(N~{\mbox{times}}),\nonumber\\
 &&\exp(3\pi i/4)~(N~{\mbox{times}}),
 \exp(-3\pi i/4)~(N~{\mbox{times}}))~,\\
 \gamma_{1,7}={\mbox{diag}}(&&\exp(\pi i/4),
 \exp(-\pi i/4), \exp(3\pi i/4), \exp(-3\pi i/4))~.
\end{eqnarray}
$\bullet$ $M=6$ ($N=(n_3-2)/6$):
\begin{eqnarray}
 \gamma_{1,3}={\mbox{diag}}(&&
 i\exp(2\pi i/3)~(N~{\mbox{times}}),
 -i\exp(2\pi i/3)~(N~{\mbox{times}}),\nonumber\\
 &&i\exp(-2\pi i/3)~(N~{\mbox{times}}),
 -i\exp(-2\pi i/3)~(N~{\mbox{times}})),\nonumber\\
 &&i~((N-1)~{\mbox{times}}),
 -i~((N-1)~{\mbox{times}}))~,\\ 
 \gamma_{1,7}={\mbox{diag}}(&&
 i\exp(2\pi i/3),-i\exp(2\pi i/3),i\exp(-2\pi i/3),-i\exp(-2\pi i/3))~.
\end{eqnarray}
The massless open string spectra of these models is given in Table \ref{Z4Z6}. 

{}Note that the one-loop $\beta$-function coefficients for the non-Abelian subgroups of the 33 sector gauge group vanish in both the ${\bf Z}_4$ and ${\bf Z}_6$ theories. In the ${\bf Z}_6$ case we have an ``accidental'' cancellation similar to that discussed in section III.

\section{Chiral ${\cal N}=0$ Gauge Theories}

{}In this section we construct four dimensional ${\cal N}=1$ gauge 
theories form Type IIB orientifolds with the $B$-field.
We start from Type IIB string theory on ${\cal M}=
{\bf C}^3/\Gamma$, where $\Gamma\approx {\bf Z}_2\otimes 
{\bf Z}_2\otimes {\bf Z}_3$, and ${\bf Z}_6$. In both cases we include the discrete torsion $T$
discussed in the previous section. This discrete torsion breaks supersymmetry completely. 
Next, we consider the $\Omega J$ orientifolds of these theories. 

\subsection{The ${\bf Z}_2\otimes {\bf Z}_2\otimes {\bf Z}_3$ Models}

{}Consider the case with $\Gamma
\approx {\bf Z}_2\otimes {\bf Z}_2\otimes {\bf Z}_3$. Let
$R_1$ and ${\widetilde R}_2=R_2 T$ be 
the generators of ${\bf Z}_2\otimes {\bf Z}_2$, where $R_i$ are the same as in section IV, 
and let $g$ be the generator of   
${\bf Z}_3\subset \Gamma$ (which was also defined in section IV).
The untwisted tadpole cancellation conditions require presence of three sets of 
D7-branes with $n_i=8/2^{2b_i}$ D7-branes in each set, where $b_i=0,1/2$ are the values of the $B$-field in the $z_i$ directions. 
The number of D3-branes is arbitrary.
The solution to the twisted tadpole cancellation conditions reads ($N=(n_3+4)/6$):
\begin{eqnarray}
 &&\gamma_{g,3}={\mbox{diag}}(\omega {\bf I}_{2N},\omega^2 {\bf I}_{2N},{\bf I}_{2N-4})~,\\
 &&\gamma_{R_i,3}=i\sigma_i\otimes {\bf I}_{3N-2}~.
\end{eqnarray}
The action on D$7_i$ Chan-Paton charges is similar. 
The spectra of these models are given in Table \ref{N01}. The non-Abelian gauge
anomaly is cancelled in these models. Also, the one-loop $\beta$-function coefficients for the $SU(N)$ and $Sp(N-2)$ subgroups of the 33 sector gauge group are given by:
\begin{eqnarray}
 && b_0 (N)=+1~,\\
 && b_0 (N-1)=-2~.
\end{eqnarray} 

\subsection{The ${\bf Z}_6$ Model}

{}Consider the case $\Gamma\approx{\bf Z}_6\approx{\bf Z}_2\otimes {\bf Z}_3$, where the generators of the ${\bf Z}_2$ and ${\bf Z}_3$ subgroups 
are ${\widetilde R}=RT$ and $g$. (Here, as before, we have 
the following action on the complex coordinates $z_i$: $Rz_1=-z_1$, $Rz_2=-z_2$, $Rz_3=z_3$, $gz_i=\omega z_i$, $\omega=\exp(2\pi i/3)$.) The $B$-field is turned on in the directions corresponding to $z_3$. 
The number of the D3-branes is arbitrary, whereas the number of D7-branes (whose locations are given by points in the $z_3$ complex plane) is 4. The twisted tadpole cancellation conditions have the following solution ($N=(n_3+4)/6$):
\begin{eqnarray}
 &&\gamma_{g,3}={\mbox{diag}}(\omega {\bf I}_{2N},\omega^2 {\bf I}_{2N},{\bf I}_{2N-4})~,\\
 &&\gamma_{R,3}={\mbox{diag}}(i,-i)\otimes {\bf I}_{3N-2}~,\\
 &&\gamma_{g,7}={\mbox{diag}}(\omega {\bf I}_{2},\omega^2 {\bf I}_{2}) ~,\\
 &&\gamma_{R,7}={\mbox{diag}}(i,-i)\otimes {\bf I}_{2}~.
\end{eqnarray}
The spectrum of this model is given in Table \ref{N02}. Note that the non-Abelian gauge anomaly cancels in this model. Also, the one-loop $\beta$-function coefficients for the $SU(N)$ and $SU(N-2)$ subgroups of the 33 sector gauge group are given by:
\begin{eqnarray}
 && b_0 (N)=+3~,\\
 && b_0 (N-1)=-6~.
\end{eqnarray}

\subsection{Comments}

{}Here the following remark is in order. As we already mentioned in section II, most of the 
$\Omega J$ orientifolds of Type IIB on 
${\bf C}^3/\Gamma$ are expected to contain states which do not possess world-sheet description. In certain cases these states decouple (upon appropriately blowing up orbifold singularities). This was shown to be the case for ${\cal N}=1$ supersymmetric orientifolds
with $\Gamma\approx{\bf Z}_2\otimes{\bf Z}_2\otimes {\bf Z}_3$ and ${\bf Z}_6$ \cite{KS2,KST,zura1}\footnote{For other related works on orientifolds, see, {\em e.g.}, \cite{other,class}.}. (These are the orbifolds considered in section IV.) Note that such non-perturbative states do not appear in the ${\cal N}=1$ orientifold of Type IIB on ${\bf C}^3/
{\bf Z}_2\otimes {\bf Z}_2$. Also, they are absent in the ${\cal N}=2$ orbifolds discussed in section III \cite{blum,KST}. Absence of such states in non-chiral ${\cal N}=0$ cases (discussed in section V) was argued in \cite{zura2}. However, there it was also pointed out that for the chiral 
${\cal N}=0$ cases of this section it is not completely clear whether such states decouple. In other cases one was able to perform various checks confirming the required decoupling. However, in the cases at hand it appears to be rather difficult to find such an independent check. In order for the gauge theories constructed in this section to have a well defined world-sheet description, it is necessary that the corresponding non-perturbative states decouple. We will {\em assume} that this is indeed the case. (Thus, non-trivial cancellation of non-Abelian gauge anomalies in these models might be considered (at least indirect) evidence for the correctness of this assumption.) Here we would like to stress that all the other gauge theories discussed in this paper are safe from such non-perturbative contributions. 

\section{Discussion}

{}The gauge theories we constructed in this paper have the property that in the large $N$ limit computation of any $M$-point correlation function reduces to the corresponding computation in the parent ${\cal N}=4$ gauge theory with the unitary gauge group. Thus, in the large $N$ limit these theories do not appear to be different from those without the $B$-flux. However, for finite $N$ the theories with and without the $B$-flux are distinct. In particular, $1/N$ corrections to the correlation functions are different in these two classes of theories. 

{}Here we would like to comment on the effect of the $B$-field on the gauge theories from the supergravity viewpoint \cite{ads}. The above orientifolds can be viewed as compactifications
of Type IIB string theory on $AdS_5\otimes X_5$, where $X_5$ is defined as follows\footnote{An example of a non-orbifold $X_5$ was discussed in \cite{KW}.}. To consider supergravity duals of the gauge theories without the orientifold planes \cite{KaSi,LNV,BKV} we consider $X_5=S^5/\Gamma$, where $\Gamma$ is the orbifold group
acting on the complex coordinates $z_i$ (where $\sum_i |z_i|^2=r^2$ defines $S^5$ with radius $r$). Here we assume that all ${\bf Z}_2$ subgroups of $\Gamma$ (if any) are such that the corresponding generators have a set of fixed points of real dimension 2 when acting on ${\bf C}^3$ (so that Type IIB on ${\bf C}^3/\Gamma$ is a modular invariant theory). To introduce duals of the orientifold planes we consider manifolds ${\widetilde X}_5=S^5/{\widetilde \Gamma}$, where ${\widetilde \Gamma}\approx\Gamma\otimes {\bf Z}_2$, and this last ${\bf Z}_2$ subgroup is generated by $J$ ($Jz_i=-z_i$). Now, suppose $\Gamma$ contains a ${\bf Z}_2$ twist $R$ which, say, has the following action: $Rz_1=-z_1, Rz_2=-z_2, Rz_3=z_3$. Then the element $RJ$ of ${\widetilde \Gamma}$ twists $z_3$ ($RJz_3=-z_3$) while leaving $z_1,z_2$ invariant. In the flat space limit this would imply that we are considering Type IIB on 
${\bf C}^2\otimes ({\bf C}/{\bf Z}_2)$ (further modded out by other elements of ${\widetilde \Gamma}$). Since $({\bf C}/{\bf Z}_2)$ is the non-compact limit of ${\bf P}^1$, it is natural to use the F-theory \cite{vafa} framework to understand the geometric origin of the effect of the $B$-flux on the corresponding $AdS_5\otimes {\widetilde X}_5$ compactification. Note that the $B$-flux is in the directions corresponding to $z_3$. This can be viewed in the context of
F-theory with quantized fluxes \cite{flux} (also see \cite{wit}). In particular, in this framework one can understand the rank reduction of the 77 open string sector gauge group geometrically\footnote{Large $N$ gauge theories without the $B$-flux in the context of F-theory
were discussed in \cite{AFM}.}.

{}Finally, we would like to mention that the large $N$ gauge theories can also be constructed
from orientifolds with orientifold 7-planes with D7- and D3-branes but without orientifold 3-planes.
Some gauge theories of this kind without the $B$-flux were studied in \cite{FS}. It would be interesting to study such theories with the $B$-flux in more detail\footnote{It would also be interesting to include non-trivial discrete torsion between various elements of the orbifold group $\Gamma$ along the lines of \cite{doug}.}.

\acknowledgments

{}This work was supported in part by the grant NSF PHY-96-02074, 
and the DOE 1994 OJI award. 
I would like to thank Gary Shiu for discussions.
I would also like to thank Albert and Ribena Yu for 
financial support.

%%%%%%%%%%%%%Table I %%%%%%%%
%%%%%%%%%%%%%%%%%%%%%%%%%%%%%%%%%%%%%%%%%%%%%%%%%%%%%%%%%%%%%%%%%%%%%%%%%%%%%%%
\begin{table}[t]
\begin{tabular}{|c|c|l|}
%%%%%%%%%%%%%%%%%%%%%%%%%%%%%%%%%%%%%%%%%%%%%%%%%%%%%%%%%%%%%%%%%%%%%%%%%%%%
 Model & Gauge Group &  Charged   \\
       &                &Hypermultiplets  \\
\hline
%%%%%%%%%%%%%%%%%%%%%%%%%%%%%%%%%%%%%%%%%%%%%%%%%%%%%%%%%%%%%%%%%%%%%%%%%%%
${\bf Z}_2$ & $U(N)_{33} \otimes U(2)_{77}$ & 
 $2 ({\bf A};{\bf 1})_{33}$  \\
               &                       & $2 ({\bf 1};{\bf 1})_{77}$ \\
               &                       & $2 ({\bf N};{\bf 2})_{37}$ \\
\hline
%%%%%%%%%%%%%%%%%%%%%%%%%%%%%%%%%%%%%%%%%%%%%%%%%%%%%%%%%%%%%%%%%%%%%%%%
${\bf Z}_4$  & $[U(N)^2]_{33}\otimes [U(1)^2]_{77}$ 
& $({\bf A},{\bf 1})_{33}$ \\ 
 & & $({\bf 1},{\bf A})_{33}$ \\
 & & $({\bf N},{\bf N})_{33}$ \\
 & & $({\bf 1},{\bf 1})_{77}$ \\
 & & $2({\bf N},{\bf 1})_{37}$ \\
 & & $2({\bf 1},{\bf N})_{37}$ \\
\hline
%%%%%%%%%%%%%%%%%%%%%%%%%%%%%%%%%%%%%%%%%%%%%%%%%%%%%%%%%%%%%%%%%%%%%%%%
${\bf Z}_6$  & $[U(N)^2 \otimes U(N-1)]_{33}\otimes  [U(1)^2]_{77}$ 
&$({\bf A},{\bf 1},{\bf 1})_{33}$ \\
 &&
$({\bf 1},{\bf A},{\bf 1})_{33}$ \\
 & & $({\bf N},{\bf 1},{\bf N-1})_{33}$ \\
 & & $({\bf 1},{\bf N},{\bf N-1})_{33}$ \\
 & & $2({\bf N},{\bf 1},{\bf 1})_{37}$ \\
 & & $2({\bf 1},{\bf N},{\bf 1})_{37}$ \\
\hline
%%%%%%%%%%%%%%%%%%%%%%%%%%%%%%%%%%%%%%%%%%%%%%%%%%%%%%%%%%%%%%%%%%%%%%%%%%
\end{tabular}
%%%%%%%%%%%%%%%%%%%%%%%%%%%%%%%%%%%%%%%%%%%%%%%%%%%%%%%%%%%%%%%%%%%%%%%%%%%
\caption{The massless open string 
spectra of ${\cal N}=2$ orientifolds of Type IIB on ${\bf C}^3/{\bf Z}_M$
$M=2,4,6$ with the $B$-field. 
The semi-colon in the column ``Charged Hypermultiplets'' separates $33$ and 
$77$ representations. The notation ${\bf A}$ stands for the two-index antisymmetric
representation (which is $N(N-1)/2$ dimensional for $U(N)$) of the corresponding unitary 
group. The $U(1)$ charges are not shown.}
\label{spectrum1} 
\end{table}
%%%%%%%%%%%%%%%%%%%%%%%%%%%%%%%%%%%%%%%%%%%%%%%%%%%%%%%%%%%%%%%%%%%%%%%%%%%%%%%

%%%%%%%%%%%%%Table II %%%%%%%%
%%%%%%%%%%%%%%%%%%%%%%%%%%%%%%%%%%%%%%%%%%%%%%%%%%%%%%%%%%%%%%%%%%%%%%%%%%%%%%%
\begin{table}[t]
\begin{tabular}{|c|c|l|}
%%%%%%%%%%%%%%%%%%%%%%%%%%%%%%%%%%%%%%%%%%%%%%%%%%%%%%%%%%%%%%%%%%%%%%%%%%%%
 Model & Gauge Group & \phantom{Hy} Charged   
  \\
       &                &Chiral Multiplets 
 \\
\hline
%%%%%%%%%%%%%%%%%%%%%%%%%%%%%%%%%%%%%%%%%%%%%%%%%%%%%%%%%%%%%%%%%%%%%%%%%%%
${\bf Z}_2\otimes{\bf Z}_2$ & $Sp(N)_{33} \otimes$  & 
 $3 \times ({\bf A})_{33}$
 \\
               &  $\bigotimes_{i=1}^3 Sp(m_i)_{7_i 7_i}$   
                  & $3 \times ({\bf a}_i)_{7_i 7_i}$   \\
                   &  & $k_i \times ({\bf N};{\bf m}_i)_{3 7_i}$ \\
                &  & $(k_i k_j) \times({\bf m}_i;{\bf m}_j)_{7_i 7_j}$ \\
\hline
%%%%%%%%%%%%%%%%%%%%%%%%%%%%%%%%%%%%%%%%%%%%%%%%%%%%%%%%%%%%%%%%%%%%%%%%%%%
${\bf Z}_2\otimes{\bf Z}_2\otimes {\bf Z}_3$ & $[U(N)\otimes Sp(N-2)]_{33} \otimes$ &
 $3 \times ({\bf A},{\bf 1})(+2)_{33}$ 
 \\
               &  $\bigotimes_{i=1}^3 U(r_i)_{7_i 7_i}$   
                  & $3 \times ({\overline {\bf N}},{\bf N-2})(-1)_{33}$  \\
           &&       $3 \times ({\bf c}_i)(+2_i)_{7_i7_i}$ \\
     & & $k_i \times ({\bf N},{\bf 1};{\bf r}_i)(+1;+1_i)_{3 7_i}$ \\
                 & & $k_i \times ({\bf 1},{\bf N-2};{\bf r}_i)(0;-1_i)_{3 7_i}$  \\
                &  & $(k_ik_j)\times({\bf r}_i;{\bf r}_j)(+1_i;+1_j)_{7_i 7_j}$  \\
\hline
%%%%%%%%%%%%%%%%%%%%%%%%%%%%%%%%%%%%%%%%%%%%%%%%%%%%%%%%%%%%%%%%%%%%%%%%%%
${\bf Z}_6$ & $[U(N)^2 \otimes U(N-2)]_{33}\otimes[U(1)^2]_{77}$  & 
  $2 \times ({\bf A},{\bf 1},{\bf 1})(+2,0,0)_{33}$ 
 \\
              & 
                      & $2 \times ({\bf 1},{\overline {\bf A}},{\bf 1})(0,-2,0)_{33}$  \\
              &    & $2 \times ({\overline {\bf N}},{\bf 1},{\overline {\bf N-2}})(-1,0,-1)_{33}$  \\
              &    & $2 \times ({\bf 1},{\bf N},{\bf N-2})(0,+1,+1)_{33}$  \\
              &    & $ ({\bf N},{\overline {\bf N}},{\bf 1})(+1,-1,0)_{33}$  \\
              &    & $ ({\overline {\bf N}},{\bf 1},{\bf N-2})(-1,0,+1)_{33}$   \\
              &    & $ ({\bf 1},{\bf N},{\overline {\bf N-2}})(0+1,-1)_{33}$  \\
              &    & $ (+1,-1)_{77}$  \\
              &    & $ 2\times ({\bf N}, {\bf 1},{\bf 1})(+1,0,0;+1,0)_{37}$  \\
              &    & $ 2\times({\bf 1},{\bf 1},{\bf N-2})(0,0,+1;0,+1)_{37}$  \\
              &    & $ 2\times({\bf 1},{\overline {\bf N}},{\bf 1})(0,-1,0;0,-1)_{37}$  \\
              &    & $ 2\times({\bf 1},{\bf 1},{\overline {\bf N-2}})(0,0,-1;-1,0)_{37}$  \\
\hline
%%%%%%%%%%%%%%%%%%%%%%%%%%%%%%%%%%%%%%%%%%%%%%%%%%%%%%%%%%%%%%%%%%%%%%%%%%
\end{tabular}
%%%%%%%%%%%%%%%%%%%%%%%%%%%%%%%%%%%%%%%%%%%%%%%%%%%%%%%%%%%%%%%%%%%%%%%%%%%
\caption{The massless open string 
spectra of the ${\cal N}=1$ orientifolds of Type IIB on ${\bf C}^3/
{\bf Z}_2\otimes {\bf Z}_2$, ${\bf C}^3/
{\bf Z}_2\otimes {\bf Z}_2\otimes {\bf Z}_3$ and ${\bf C}^3/
{\bf Z}_6$ with the $B$-field.
The notation ${\bf A}$ for the $Sp$ groups stands for the two-index antisymmetric (reducible)
representation. Also, note that we are using the normalization where the rank of $Sp(2K)$ is $K$.
The $U(1)$ charges are given in parentheses. We have introduced the following notations:
$k_i=2^{2b_i}$, $m_i=4/k_i$, $a_i=m_i(m_i-1)/2$, $r_i=2/k_i$, $c_i=r_i(r_i-1)/2$. (For $c_i=0$ the corresponding state is absent.)}
\label{Z2} 
\end{table}
%%%%%%%%%%%%%%%%%%%%%%%%%%%%%%%%%%%%%%%%%%%%%%%%%%%%%%%%%%%%%%%%%%%%%%%%%%%%%%%

%%%%%%%%%%%%%Table III %%%%%%%%
%%%%%%%%%%%%%%%%%%%%%%%%%%%%%%%%%%%%%%%%%%%%%%%%%%%%%%%%%%%%%%%%%%%%%%%%%%%%%%%
\begin{table}[t]
\begin{tabular}{|c|c|c|l|}
%%%%%%%%%%%%%%%%%%%%%%%%%%%%%%%%%%%%%%%%%%%%%%%%%%%%%%%%%%%%%%%%%%%%%%%%%%%%
 Model &Gauge Group  & Charged Bosons &Charged Fermions
  \\
 \hline
%%%%%%%%%%%%%%%%%%%%%%%%%%%%%%%%%%%%%%%%%%%%%%%%%%%%%%%%%%%%%%%%%%%%%%%%%%%
&&&\\
 ${\bf Z}_4$ & $[U(N)^2]_{33}\otimes [U(1)^2]_{77}$  &  
 $[({\bf 1},{\bf 1})_b]_{33}$ &
 
 \\ 
&   &  
 ${1\over 2}[({\bf Adj},{\bf 1})_b]_{33}$ &
 $[({\bf N},{\overline {\bf N}})_f]_{33}$
 \\
 &   &  
 ${1\over 2}[({\bf 1},{\bf Adj})_b]_{33}$ &
 $[({\overline {\bf N}},{\bf N})_f]_{33}$
 \\
 &   &  
 $[({\bf A},{\bf 1})_b]_{33}$ &
 $[({ {\bf A}},{\bf 1})_f]_{33}$
 \\
 &   &  
 $[({\bf 1},{\bf A})_b]_{33}$ &
 $[({ {\bf 1}},{\bf A})_f]_{33}$
 \\
 &   &  
 $[({\bf N},{\bf N})_b]_{33}$ &
 $[({ {\bf N}},{\bf N})_f]_{33}$
 \\
 &   &  
 $2\times [({\bf 1},{\bf 1})_b]_{77}$ &
 $3\times [({ {\bf 1}},{\bf 1})_f]_{77}$
  \\
&   &  
 $2\times [({\bf N},{\bf 1})_b]_{37}$ &
 $2\times [({ {\bf N}},{\bf 1})_f]_{37}$
 \\
&   &  
 $2\times [({\bf 1},{\bf N})_b]_{37}$ &
 $2\times [({ {\bf 1}},{\bf N})_f]_{37}$
 \\
\hline
%%%%%%%%%%%%%%%%%%%%%%%%%%%%%%%%%%%%%%%%%%%%%%%%%%%%%%%%%%%%%%%%%%%%%%%%%%%
&&&\\
${\bf Z}_6$ & $[U(N)^2\otimes U(N-1)]_{33}\otimes [U(1)^2]_{77}$  &  
 $3\times {1\over 2} [({\bf 1},{\bf 1},{\bf 1})_b]_{33}$ &
 $[({\bf N},{\overline {\bf N}},{\bf 1})_f]_{33}$
 \\
            &   
 &
 ${1\over 2} [({\bf Adj},{{\bf 1}},{\bf 1})_b]_{33}$ & 
 $[({\overline {\bf N}},{\bf N},{\bf 1})_f]_{33}$\\
           &  & ${1\over 2} [({{\bf 1}},{\bf Adj},{\bf 1})_b]_{33}$ &
 $[({ {\bf 1}},{\bf 1},{\bf A})_f]_{33}$ \\
           &  & ${1\over 2} [({{\bf 1}},{\bf 1},{\bf Adj})_b]_{33}$ &
 $[({ {\bf 1}},{\bf 1},{\overline {\bf A}})_f]_{33}$ \\
            &   &
 $[({\bf A},{{\bf 1}},{\bf 1})_b]_{33}$ & 
     $[({ {\bf N}},{\bf N},{ {\bf 1}})_f]_{33}$\\
           &  &  $[({{\bf 1}},{\bf A},{\bf 1})_b]_{33}$ &
 $[({ {\bf N}},{\bf 1},{ \overline {\bf N-1}})_f]_{33}$\\
            &   &
 $[({\bf N},{{\bf 1}},{\bf N-1})_b]_{33}$ &
  $[({ {\bf 1}},{\bf N},{{\bf N-1}})_f]_{33}$ \\
           &  &  $[({{\bf 1}},{\bf N},{\overline {\bf N-1}})_b]_{33}$ & \\
 &   &  
 $[({\bf 1},{\bf 1},{\bf 1})_b]_{77}$ &
 $3\times [({\bf 1},{\bf 1},{\bf 1})_f]_{77}$
 \\
 &   &  
 $2\times {1\over 2} [({\bf N},{\bf 1},{\bf 1})_b]_{37}$ & 
 $2\times [({\bf N},{\bf 1},{\bf 1})_f]_{37}$
 \\
&   &  
 $2\times {1\over 2} [({\bf 1},{\bf N},{\bf 1})_b]_{37}$ &
 $2\times [({\bf 1},{\bf N},{\bf 1})_f]_{37}$
  \\
&   &  
 $2\times [({\bf 1},{\bf 1},{\bf N-1})_b]_{37}$ &
 \\ 
\hline
%%%%%%%%%%%%%%%%%%%%%%%%%%%%%%%%%%%%%%%%%%%%%%%%%%%%%%%%%%%%%%%%%%%%%%%%%%%
\end{tabular}
%%%%%%%%%%%%%%%%%%%%%%%%%%%%%%%%%%%%%%%%%%%%%%%%%%%%%%%%%%%%%%%%%%%%%%%%%%%
\caption{The massless open string spectra of the ${\cal N}=0$ orientifolds of Type IIB on ${\bf C}\otimes ({\bf C}^2/{\bf Z}_4)$ and ${\bf C}\otimes ({\bf C}^2/{\bf Z}_6)$
with the $B$-field. The subscript ``$b$'' indicates that the corresponding field consists of the bosonic content of a hypermultiplet. (Thus, ${1\over 2}$ of this content corresponds to a complex scalar.)
The subscript ``$f$'' indicates that the corresponding field consists of the fermionic content
of a hypermultiplet (that is, of one left-handed and one right-handed chiral fermion in the corresponding representation of the gauge group). The notation ${\bf Adj}$ stands for the adjoint representation (not including the $U(1)$ factor). The $U(1)$ charges are not shown.}
\label{Z4Z6} 
\end{table}
%%%%%%%%%%%%%%%%%%%%%%%%%%%%%%%%%%%%%%%%%%%%%%%%%%%%%%%%%%%%%%%%%%%%%%%%%%%%%%%

%%%%%%%%%%%%%Table IV%%%%%%%%
%%%%%%%%%%%%%%%%%%%%%%%%%%%%%%%%%%%%%%%%%%%%%%%%%%%%%%%%%%%%%%%%%%%%%%%%%%%%%%%
\begin{table}[t]
\begin{tabular}{|c|c|l|}
%%%%%%%%%%%%%%%%%%%%%%%%%%%%%%%%%%%%%%%%%%%%%%%%%%%%%%%%%%%%%%%%%%%%%%%%%%%%
 Model &Gauge Group   & Charged Complex Bosons 
  \\
 \hline
%%%%%%%%%%%%%%%%%%%%%%%%%%%%
%%%%%%%%%%%%%%%%%%%%%%%%%%%%%%%%%%%%
${\bf Z}_2\otimes{\bf Z}_2\otimes {\bf Z}_3$, $T$ & $[U(N)\otimes Sp(N-2)]_{33}\otimes$  &  
 
 $3\times [({\bf A},{\bf 1})(+2)_c]_{33}$
 \\
           &  $\bigotimes_{i=1}^3 [U(r_i)]_{7_i 7_i}$ 
 &  
 $3\times [({\overline {\bf N}},{{\bf N-2}})(-1)_c]_{33}$  \\
   &   &  
 
 $3\times [({\bf 1}_i)(+r_i)_c]_{7_i7_i}$
 \\
            &   
 &  
 $k_1\times [({\bf N},{\bf 1};{\bf r}_1)(+1;+1_1)_c]_{37_1}$  \\
            &   
 &  
 $k_1\times [({\bf 1},{{\bf N-2}};{{\bf r}}_1)(0;-1_1)_c]_{37_1}$  \\
 &&  
 $(k_2 k_3)\times [({\bf r}_2;{\bf r}_{3}) (+1_2;+1_{3})_c]_{7_2
 7_{3}}$  \\
            &   
 &  
 $k_s \times [({\bf N},{\bf 1};{{\bf r}}_s)(+1;-1_s)_c]_{37_s}$  \\
          &   
 &  
 $k_s \times [({\overline {\bf N}},{\bf 1};{\bf r}_s)(-1;+1_s)_c]_{37_s}$  \\
          &   
 & 
 $(k_1 k_s)\times [({\bf r}_1;{{\bf r}}_s)(+1_1;-1_s)_c]_{7_1 7_s}$  \\
\hline
%%%%%%%%%%%%%%%%%%%%%%%%%%%%%%%%%%%%%%%%%%%%%%%%%%%%%%%%%%%%%%%%%%%%%%%%%%%%
 &&  Charged Chiral Fermions 
  \\
 \hline
%%%%%%%%%%%%%%%%%%%%%%%%%%%%
%%%%%%%%%%%%%%%%%%%%%%%%%%%%%%%%%%%%
 &  &  
 $ [({\bf Adj},{\bf 1})(0)_L]_{33}$
 \\ 
            &   
 &
 $ [({\bf 1},{\bf a})(0)_L]_{33}$
 \\
 &&
 $2\times [({\bf 1},{\bf 1})(0)_L]_{33}$
 \\
  &   &  
 $ [({\bf S},{\bf 1})(+2)_L]_{33}$
 \\
&   &  
 $2\times [({\bf A},{\bf 1})(+2)_L]_{33}$
 \\
            &   
 & 
 $3\times [({\overline {\bf N}},{{\bf N-2}})(-1)_L]_{33}$  \\
  &   &  
 $ [({\bf p}_i)(+2_i)_L]_{7_i7_i}$
 \\
  &   &  
 $2\times [({\bf c}_i)(+2_i)_L]_{7_i7_i}$
 \\
 &   &  
 $ [({\bf q}_i)(0_i)_L]_{7_i 7_i}$
 \\ 
 & &
 $2\times [({\bf 1}_i)(0)_L]_{7_i 7_i}$
 \\
            &   
 &  
 $k_1\times [({\bf N},{\bf 1};{\bf r}_1)(+1;+1_1)_L]_{37_1}$  \\
            &   
 &  
 $k_1\times [({\bf 1},{{\bf N-2}};{{\bf r}}_1)(0;-1_1)_L]_{37_1}$  \\
 &&  
 $(k_2 k_3)\times [({\bf r}_2,{\bf 1}_2;{\bf r}_{3},{{\bf 1}}_{3}) (+1_2;+1_{3})_L]_{7_2
 7_{3}}$  \\
            &   
 &  
 $k_s \times [({\overline {\bf N}},{\bf 1};{{\bf r}}_s)(-1;-1_s)_L]_{37_s}$  \\
            &   
 &  
 $k_s \times [({\bf 1},{{\bf N-2}};{{\bf r}}_s)(0;+1_s)_L]_{37_s}$  \\
            &   
 &  
 $(k_1 k_s)\times [({\bf r}_1,{\bf r}_s)(-1_1;-1_s)_L]_{7_1 7_s}$  \\
 \hline
%%%%%%%%%%%%%%%%%%%%%%%%%%%%%%%%%%%%%%%%%%%%%%%%%%%%%%%%%%%%%%%%%%%%%%%%%%%
\end{tabular}
%%%%%%%%%%%%%%%%%%%%%%%%%%%%%%%%%%%%%%%%%%%%%%%%%%%%%%%%%%%%%%%%%%%%%%%%%%%
\caption{The massless open string spectrum of the ${\cal N}=0$ orientifold of Type IIB on ${\bf C}^3/{\bf Z}_2\otimes{\bf Z}_2\otimes {\bf Z}_3$ with the $B$-field.
The subscript ``$c$'' indicates that the corresponding field is a complex boson.
The subscript ``$L$'' indicates that the corresponding field is a left-handed chiral
fermion. The notation ${\bf S}$ stands for the two-index symmetric representation of $SU(N)$. 
The notation ${\bf Adj}$ stands for the $N^2-1$ dimensional adjoint representation of $SU(N)$,
whereas ${\bf a}$ stands for the $N(N-1)/2-1$ dimensional {\em traceless} antisymmetric representation of $Sp(N)$.
Also, $s=2,3$, and $k_i,r_i,c_i$ were defined in Table II. Here we have introduced the following notation: $p_i=r_i(r_i+1)/2$, $q_i=r^2_i-1$. (For $q_i=0$ the corresponding state is absent.)
The $U(1)$ charges are
given in parentheses.}
\label{N01} 
\end{table}
%%%%%%%%%%%%%%%%%%%%%%%%%%%%%%%%%%%%%%%%%%%%%%%%%%%%%%%%%%%%%%%%%%%%%%%%%%%

%%%%%%%%%%%%%Table V %%%%%%%%
%%%%%%%%%%%%%%%%%%%%%%%%%%%%%%%%%%%%%%%%%%%%%%%%%%%%%%%%%%%%%%%%%%%%%%%%%%%%%%%
\begin{table}[t]
\begin{tabular}{|c|c|l|}
%%%%%%%%%%%%%%%%%%%%%%%%%%%%%%%%%%%%%%%%%%%%%%%%%%%%%%%%%%%%%%%%%%%%%%%%%%%%
 Model &Gauge Group   & Charged Complex Bosons 
  \\
 \hline
%%%%%%%%%%%%%%%%%%%%%%%%%%%%%%%%%%%%%%%%%%%%%%%%%%%%%%%%%%%%%%%%%%%%%%%%%%%
${\bf Z}_6$, $T$ & $[U(N)^2\otimes U(N-2)]_{33}\otimes [U(1)^2]_{77}$  &  
 $2\times [({\bf A},{\bf 1},{\bf 1})(+2,0,0)_c]_{33}$
 \\
            &   
  &
 $2\times [({\bf 1},{\overline {\bf A}},{\bf 1})(0,-2,0)_c]_{33}$  \\
            & 
 &
 $2\times [({\overline {\bf N}},{\bf 1},{\overline {\bf N-2}})(-1,0,-1)_c]_{33}$  \\
         & 
 &
 $2\times [({\bf 1},{\bf N},{\bf N-2})(0,+1,+1)_c]_{33}$  \\
 & &
  $[({\overline {\bf N}},{\bf 1},{\bf N-2})(-1,0,+1)_c]_{33}$  \\
          & 
 &
 $[({\bf 1},{\bf N},{\overline {\bf N-2}})(0,+1,-1)_c]_{33}$  \\
 & &$ [({\bf N},{\overline {\bf N}},{\bf 1})(+1,-1,0)_c]_{33}$  \\
 & &$  [(+1,-1)_c]_{77}$  \\
&         & $2\times [({\bf N},{\bf 1},{\bf 1})
(+1,0,0;0,+1)_c]_{37}$\\
&         & $2\times [({\bf 1},{\bf N},{\bf 1})
(0,+1,0;+1,0)_c]_{37}$\\
&         & $2\times [({\overline{\bf N}},{\bf 1},{\bf 1})
(-1,0,0;0,-1)_c]_{37}$\\
&         & $2\times [({\bf 1},{\overline {\bf N}},{\bf 1})
(0,-1,0;-1,0)_c]_{37}$\\
\hline
%%%%%%%%%%%%%%%%%%%%%%%%%%%%%%%%%%%%%%%%%%%%%%%%%%%%%%%%%%%%%%%%%%%%%%%%%%%
 & & Charged Chiral Fermions 
  \\
 \hline
%%%%%%%%%%%%%%%%%%%%%%%%%%%%%%%%%%%%%%%%%%%%%%%%%%%%%%%%%%%%%%%%%%%%%%%%%%%
 &   &
 $[({\bf A},{\bf 1},{\bf 1})(+2,0,0)_L]_{33}$
 \\
            &   
 &
 $[({\bf 1},{\overline {\bf A}},{\bf 1})(0,-2,0)_L]_{33}$  \\
            & 
 &
 $[({\overline {\bf N}},{\bf 1},{\overline {\bf N-2}})(-1,0,-1)_L]_{33}$  \\
          & 
 &
 $[({\bf 1},{\bf N},{\bf N-2})(0,+1,+1)_L]_{33}$  \\
 & &
  $2\times [({\overline {\bf N}},{\bf 1},{\bf N-2})(-1,0,+1)_L]_{33}$  \\
          & 
 &  
 $2\times [({\bf 1},{\bf N},{\overline {\bf N-2}})(0,+1,-1)_L]_{33}$  \\
 & &$2\times  [({\bf N},{\overline {\bf N}},{\bf 1})(+1,-1,0)_L]_{33}$  \\
  & &$ [({\bf N},{{\bf N}},{\bf 1})(+1,+1,0)_L]_{33}$  \\
 & &$ [({\overline {\bf N}},{\overline {\bf N}},{\bf 1})(-1,-1,0)_L]_{33}$  \\ 
   & &$ [({\bf 1},{{\bf 1}},{\bf A})(0,0,+2)_L]_{33}$  \\
 & &$ [({{\bf 1}},{{\bf 1}},{\overline {\bf A}})(0,0,-2)_L]_{33}$  \\
 & &$2\times   [(+1,-1)_L]_{77}$  \\
  & &$ [(+1,+1)_L]_{77}$  \\
 & &$ [(-1,-1)_L]_{77}$  \\     
&         & $2\times [({\overline {\bf N}},{\bf 1},{\bf 1})
(-1,0,0;-1,0)_L]_{37}$\\
& & $2\times [({\bf 1},{\bf 1},{\overline {\bf N-2}})
(0,0,-1;0,-1)_L]_{37}$\\
 &         & $2\times [({\bf 1},{\bf N},{\bf 1})(0,+1,0;0,+1)_L]_{37}$  \\ 
 & & $2\times [({\bf 1},{\bf 1},{{\bf N-2}})
(0,0,+1;+1,0)_L]_{37}$\\
%%%%%%%%%%%%%%%%%%%%%%%%%%%%%%%%%%%%%%%%%%%%%%%%%%%%%%%%%%%%%%%%%%%%%%%%%%%
\end{tabular}
%%%%%%%%%%%%%%%%%%%%%%%%%%%%%%%%%%%%%%%%%%%%%%%%%%%%%%%%%%%%%%%%%%%%%%%%%%%
\caption{The massless open string spectrum of the ${\cal N}=0$ orientifold of Type IIB on ${\bf C}^3/
{\bf Z}_6$ with the $B$-field. The subscript ``$c$'' indicates that the corresponding field is a complex boson.
The subscript ``$L$'' indicates that the corresponding field is a left-handed chiral
fermion. The $U(1)$ charges are
given in parentheses.}
\label{N02} 
\end{table}
%%%%%%%%%%%%%%%%%%%%%%%%%%%%%%%%%%%%%%%%%%%%%%%%%%%%%%%%%%%%%%%%%%%%%%%%%%%%%%%


\begin{references}

\bibitem{ads} See, {\em e.g.},\\
I.R. Klebanov, ``World Volume Approach to Absorption by
Non-dilatonic Branes'', Nucl. Phys. {\bf B496} (1997) 231, hep-th/9702076;\\
S.S. Gubser and I.R. Klebanov, ``Absorption by Branes and Schwinger 
Terms in the World Volume Theory'', Phys. Lett. {\bf B413} (1997) 41,
hep-th/9708005;\\
J.M. Maldacena, ``The Large $N$ Limit of
Superconformal Field
Theories and Supergravity'', hep-th/9711200;\\
S.S.  Gubser, I.R. Klebanov and A.M. Polyakov, ``Gauge Theory 
Correlators from
Non-Critical String Theory'', hep-th/9802109;\\
E. Witten, ``Anti-de Sitter
Space And Holography'', hep-th/9802150; ``Anti-de Sitter Space, Thermal Phase Transition, And Confinement In Gauge Theories'', hep-th/9803131.

\bibitem{KaSi} S. Kachru and E. Silverstein, ``4d Conformal Field Theories
and Strings on Orbifolds'', Phys. Rev. Lett. {\bf 80} (1998) 4855, hep-th/9802183.

\bibitem{LNV} A. Lawrence, N. Nekrasov and C. Vafa, ``On Conformal Theories in 
Four Dimensions'', hep-th/9803015.

\bibitem{BKV} M. Bershadsky, Z. Kakushadze and C. Vafa, 
``String Expansion as Large $N$ Expansion of Gauge Theories'', Nucl. Phys. {\bf B523} (1998) 59, hep-th/9803076.

\bibitem{thooft} G. 't Hooft, ``A Planar Diagram Theory For Strong
Interactions'', Nucl. Phys. {\bf B72} (1974) 461.

\bibitem{CS} E.Witten, ``Chern-Simons Gauge Theory As A String Theory'', hep-th/9207094.

\bibitem{zura} Z. Kakushadze, ``Gauge Theories from Orientifolds and Large $N$ Limit'', 
Nucl. Phys. {\bf B529} (1998) 157, hep-th/9803214.

\bibitem{BJ} M. Bershadsky and A. Johansen, ``Large $N$ limit of {\em orbifold} field 
theories'', hep-th/9803249.

\bibitem{KST} Z. Kakushadze, G. Shiu and S.-H.H. Tye, ``Type IIB Orientifolds,
F-theory, Type I Strings on Orbifolds and Type I - Heterotic Duality'', 
hep-th/9804092.

\bibitem{zura1} Z. Kakushadze, ``On Large $N$ Gauge Theories from Orientifolds'',
hep-th/9804184.

\bibitem{zura2} Z. Kakushadze, ``Anomaly Free Non-Supersymmetric Gauge Theories From Orientifolds'', hep-th/9806091 

\bibitem{rev} Z. Kakushadze, ``String Expansion as 't Hooft's Expansion'', hep-th/9808019.

\bibitem{class} Z. Kakushadze, ``On Four Dimensional ${\cal N}=1$ Type I Compactifications'',
hep-th/9806008.

\bibitem{Bij} M. Bianchi, G. Pradisi and A. Sagnotti, ``Toroidal Compactifications and Symmetry Breaking in Open String Theories'', Nucl. Phys. {\bf B376} (1992) 365;\\
M. Bianchi, ``A Note on Toroidal Compactifications of the Type I Superstring and Other Superstring Vacuum Configurations with 16 Supercharges'', hep-th/9711201.

\bibitem{wit} E. Witen, ``Toroidal Compactification Without Vector Structure'', J. High Energy Phys. {\bf 02} (1998) 006, hep-th/9712028.

\bibitem{NS} Z. Kakushadze, G. Shiu and S.-H.H. Tye, 
``Type IIB Orientifolds
with NS-NS Antisymmetric Tensor Backgrounds'', Phys. Rev. {\bf D58}
(1998) 086001, hep-th/9803141.

\bibitem{KS2}Z. Kakushadze, ``Aspects of $N=1$ Type I-Heterotic Duality in Four 
Dimensions'', Nucl. Phys. {\bf B512} (1998) 221, hep-th/9704059;\\ 
Z. Kakushadze and G. Shiu, ``4D Chiral $N=1$ Type I Vacua with and 
without D5-branes'', Nucl. Phys. {\bf B520} (1998) 75,
hep-th/9706051.

\bibitem{other} G. Pradisi and A. Sagnotti, ``Open String Orbifolds'',
Phys. Lett. {\bf B216} (1989) 59;\\
M. Bianchi and A. Sagnotti, ``On the Systematics of Open String Theories'',
Phys. Lett. {\bf B247} (1990) 517; ``Twist Simmetry and Open String Wilson Lines'',
Nucl. Phys. {\bf B361} (1991) 519;\\
E.G. Gimon and J. Polchinski, ``Consistency Conditions 
for Orientifolds and D-Manifolds'', Phys. Rev. {\bf D54} (1996) 1667, hep-th/9601038;\\
E.G. Gimon and C.V. Johnson, ``K3 Orientifolds'', Nucl. Phys. {\bf B477} (1996) 715, hep-th/9604129;\\
A. Dabholkar and J. Park, ``Strings on Orientifolds'', Nucl. Phys. {\bf B477} (1996) 701, 
hep-th/9604178;\\
M. Berkooz and R.G. Leigh, ``A D=4 N=1 Orbifold of Type I Strings'',
Nucl. Phys. {\bf B483} (1997) 187,
hep-th/9605049;\\
C. Angelantonj, M. Bianchi, G. Pradisi, A. Sagnotti and 
Ya.S. Stanev, ``Chiral Asymmetry in Four-Dimensional Open-String Vacua'', 
Phys. Lett. {\bf B385} (1996) 96, hep-th/9606169;\\
Z. Kakushadze and G. Shiu, ``A Chiral $N=1$ Type I Vacuum in Four Dimensions and Its Heterotic Dual'', Phys. Rev. {\bf D56} (1997) 3686,
hep-th/9705163;\\
G. Zwart, ``Four-dimensional $N=1$ $Z_N \times Z_M$ Orientifolds'', hep-th/9708040;\\
G. Aldazabal, A. Font, L.E. Ib{\'a}{\~n}ez and G. Violero, ``D=4, N=1, Type IIB Orientifolds'', hep-th/9804026.;\\
Z. Kakushadze, ``A Three-Family $SU(6)$ Type I Compactification'', Phys. Lett. {\bf B434} (1998) 269, hep-th/9804110;
``A Three-Family $SU(4)_c\otimes SU(2)_w\otimes U(1)$ 
Type I Vacuum'', hep-th/9806044;\\
G. Shiu and S.-H.H. Tye, ``TeV Scale Superstring and Extra Dimensions'', hep-th/9805157;\\
J.D. Blum, ``Anomaly Inflow at Singularities'', hep-th/9806012;\\
R. Blumenhagen and A. Wisskirchen, ``Spectra of 4D, N=1 Type I String Vacua on Non-Toroidal CY Threefolds'', hep-th/9806131;\\
Z. Kakushadze and S.-H.H. Tye, ``Three Generations in Type I Compactifications'', hep-th/9806143.

\bibitem{blum} J. Polchinski, ``Tensors from K3 Orientifolds'', Phys. Rev. {\bf D55} (1997) 6423, hep-th/9606165;\\
E.G. Gimon and C.V. Johnson, ``Multiple Realisations of N=1 Vacua in Six Dimensions'', Nucl. Phys. {\bf B479} (1996) 285, hep-th/9606176;\\
J.D. Blum, ``F Theory Orientifolds, M Theory Orientifolds, and Twisted Strings'', 
Nucl. Phys. {\bf B486} (1997) 34, hep-th/9608053;\\
P. Berglund and E.G. Gimon, ``On the Complementarity of F-theory, Orientifolds, and Heterotic Strings'', Nucl. Phys. {\bf B525} (1998) 73, hep-th/9803168.

\bibitem{KW} I.R. Klebanov and E. Witten, ``Superconformal Field Theory on Threebranes at a Calabi-Yau Singularity'', hep-th/9807080.

\bibitem{vafa} C. Vafa, ``Evidence for F-theory'', Nucl. Phys. {\bf B469} (1996) 403,
hep-th/9602022.

\bibitem{flux} M. Bershadsky, T. Pantev and V. Sadov, ``F-Theory with Quantized Fluxes'',
hep-th/9805056. 

\bibitem{AFM} O. Aharony, A. Fayyazuddin and J. Maldacena, ``The Large N Limit of ${\cal N} =2,1 $ Field Theories from Threebranes in F-theory'', hep-th/9806159.

\bibitem{FS} A. Fayyazuddin and M. Spalinski, ``Large N Superconformal Gauge Theories and Supergravity Orientifolds'', hep-th/9805096.

\bibitem{doug} M.R. Douglas, ``D-branes and Discrete Torsion'', hep-th/9807235.

\end{references}
\end{document}